\documentclass[iop]{emulateapj}

\usepackage{graphicx}
\usepackage{epstopdf}

\newcommand{\fig}[1]{Fig.~\ref{#1}}

\newcommand{\hone}{\ion{H}{1}}

\newcommand{\heone}{\ion{He}{1}}

\newcommand{\cone}{\ion{C}{1}}
\newcommand{\ctwo}{\ion{C}{2}}

\newcommand{\none}{\ion{N}{1}}
\newcommand{\ntwo}{\ion{N}{2}}

\newcommand{\oone}{\ion{O}{1}}
\newcommand{\otwo}{\ion{O}{2}}

\newcommand{\neone}{\ion{Ne}{1}}
\newcommand{\netwo}{\ion{Ne}{2}}

\newcommand{\mgone}{\ion{Mg}{1}}
\newcommand{\mgtwo}{\ion{Mg}{2}}
\newcommand{\alone}{\ion{Al}{1}}
\newcommand{\altwo}{\ion{Al}{2}}
\newcommand{\sione}{\ion{Si}{1}}
\newcommand{\sitwo}{\ion{Si}{2}}

\newcommand{\sone}{\ion{S}{1}}
\newcommand{\stwo}{\ion{S}{2}}

\newcommand{\feone}{\ion{Fe}{1}}
\newcommand{\fetwo}{\ion{Fe}{2}}
\newcommand{\fethree}{\ion{Fe}{3}}

\newcommand{\ebv}{$E($\bv)}

\newcommand{\flux}{erg cm$^{-2}$ s$^{-1}$  \AA$^{-1}$}

\newcommand{\nho}{$N$(\hone)}

\newcommand{\fuse}{{\it FUSE}}
\newcommand{\hst}{{\it HST}}

\newcommand{\iue}{{\it IUE}}

\slugcomment{To Appear in {\rm The Astrophysical Journal Letters}}

\shorttitle{Nitrogen Opacity in the Bright Star in 47 Tuc}
\shortauthors{Dixon \& Chayer}

\begin{document}


\title{Resonances in the Photoionization Cross Sections of Atomic Nitrogen Shape\\
the Far-Ultraviolet Spectrum of the Bright Star in 47 Tucanae}


\author{William V. Dixon and Pierre Chayer}
\affil{Space Telescope Science Institute, 3700 San Martin Drive, Baltimore, MD 21218}
%
%

\begin{abstract}

The far-ultraviolet (FUV) spectrum of the Bright Star (B8 III) in 47 Tuc (NGC~104) shows a remarkable pattern: 
it is well fit by LTE models at wavelengths longer than Lyman $\beta$, but at shorter wavelengths it is fainter than the models by a factor of two.  A spectrum of this star obtained with the {\em Far Ultraviolet Spectroscopic Explorer (FUSE)}\/ shows broad absorption troughs with sharp edges at 995 and 1010 \AA\ and a deep absorption feature at 1072 \AA, none of which are predicted by the models.  We find that these features are caused by resonances in the photoionization cross sections of the first and second excited states of atomic nitrogen (2s$^2$ 2p$^3$ $^2$D$^0$ and $^2$P$^0$).
Using cross sections from the Opacity Project, we can reproduce these features, but only if we use the cross sections at their full resolution, rather than the resonance-averaged cross sections usually employed to model stellar atmospheres.  
These resonances are strongest in stellar atmospheres with enhanced nitrogen and depleted carbon abundances, a  pattern typical of post-AGB stars.

\end{abstract}


\section{Introduction}

The Bright Star in the globular cluster 47 Tuc (NGC~104) is the cluster's brightest member at both optical \citep{Feast:Thackeray:60} and ultraviolet \citep{deBoer:85} wavelengths.  This blue giant (B8 III) is a post-asymptotic giant-branch (post-AGB) star: having ascended the AGB and ejected its outer envelope, it is moving across the color-magnitude diagram toward the tip of the white-dwarf cooling sequence.  \citet{Dixon:95} observed the star with the Hopkins Ultraviolet Telescope \citep[HUT;][]{HUT1CAL1}.  They found that, while a \citet{Kurucz:92} model adequately reproduces the stellar spectrum across most of the HUT bandpass (900 -- 1850 \AA), it vastly overpredicts the stellar flux at wavelengths shorter than Lyman $\beta$ (1026 \AA).  To determine the source of this opacity, we observed the star with the {\em Far Ultraviolet Spectroscopic Explorer (FUSE;} \citealt{Moos:00, Sahnow:00}).  With a spectral resolution of 0.05 \AA, \fuse\/ can resolve features that are blended at the 3 \AA\ resolution of HUT.  By fitting synthetic spectra to the star's \fuse\/ spectrum, we have determined that the peculiar structures in the stellar continuum are sculpted by resonances in the photoionization cross sections of excited-state neutral nitrogen in the stellar atmosphere.

\section{Observations and Data Reduction}\label{sec_observations}

The Bright Star in 47 Tuc was observed through the 30\arcsec\ $\times$ 30\arcsec\ low-resolution (LWRS) aperture of \fuse\/  for 12,324 s on 2006 June 07 (data set G0730101).  Archival NUV images of 47 Tuc obtained with the Ultraviolet Imaging Telescope \citep{Stecher:97} confirm that no other UV-bright stars fell within the spectrograph aperture.  The data were reduced using version 3.2 of the CalFUSE calibration pipeline \citep{Dixon:07}.  For each channel, the spectra from individual exposures were shifted to a common wavelength scale, weighted by their exposure time, and averaged.  The spectra were binned by four detector pixels, or about 0.025~\AA, approximately half of the instrument resolution.  The complete spectrum was spliced together from the available channels.  The intensities of the HUT and \fuse\/ spectra differ by 4\% in the region of overlap, consistent with the uncertainties in the flux calibration of each instrument.  For the figures in this paper, the spectra were smoothed by an additional three spectral bins.

\section{Spectral Analysis} \label{sec_analysis}

\subsection{Synthetic Spectra and Model Fitting}

We employ the LTE model atmospheres of \citet{Castelli:Kurucz:03} and the synthetic-spectral code SYNSPEC \citep{Hubeny:88} to model the Bright Star's \fuse\/ spectrum.  We adopt the atmospheric parameters derived from the HUT spectrum ($T_{\rm eff}$ = 11,000 K; $\log g$ = 2.5; [M/H] = -1.0), but use the Castelli \& Kurucz $\alpha$-enhanced models, for which the abundances of O, He, Mg, Si, S, Ar, Ca, and Ti are enhanced by +0.4 dex over the scaled-solar values, consistent with the measured abundances of giant stars in 47 Tuc \citep{Alves-Brito:05}.  These models assume a plane-parallel, homogeneous, and stationary stellar atmosphere in hydrostatic and radiative equilibrium.  Given a model atmosphere and a line list,  SYNSPEC solves the radiative transfer equation and computes a synthetic stellar spectrum.  All models assume local thermodynamic equilibrium (LTE).  It is possible to modify the atmospheric abundances within SYNSPEC, but for simplicity we adopt the values used to generate the model  atmosphere.
 
Within SYNSPEC, atomic species may be treated either explicitly or implicitly.  Explicit atoms contribute to the continuum opacity through bound-free transitions from all specified energy levels and free-free transitions from all specified ions.  Implicit atoms do not contribute to the continuum opacity, but are included in the list of bound-bound transitions.  In our model, 22 ions of 11 atoms are treated explicitly: H$^-$, \hone, \heone, \cone, \ctwo, \none, \ntwo, \oone, \otwo, \neone, \netwo, \mgone, \mgtwo, \alone, \altwo, \sione, \sitwo, \sone, \stwo, \feone, \fetwo, and \fethree.  For each explicit ion, a data file must be supplied to the program.  We employ the data files developed for B-type stars by \citet{Lanz:Hubeny:2007}.  
 
We run SYNSPEC from within the IDL program SYNPLOT, written by I.\ Hubeny,  which we have modified to include dust extinction and absorption by the interstellar medium (ISM).  We adopt an extinction \ebv\ = 0.04 \citep[2010 edition]{Harris:96} and an interstellar hydrogen column density $\log$ \nho\ = 20.62 \citep{Heiles:79}.  Other ISM features are fit to absorption lines in the \fuse\/ spectrum.  Between 1400 and 1800~\AA, the star's HUT spectrum is well fit by a Kurucz model and free of strong airglow lines \citep{Dixon:95}.  Its mean flux is $4.3 \times 10^{-13}$ \flux.  We scale the synthetic spectrum to match the observed flux in this region.  This scale factor is applied to all of the models discussed below.

\begin{figure*}
\centering
\epsscale{1.05}
\plotone{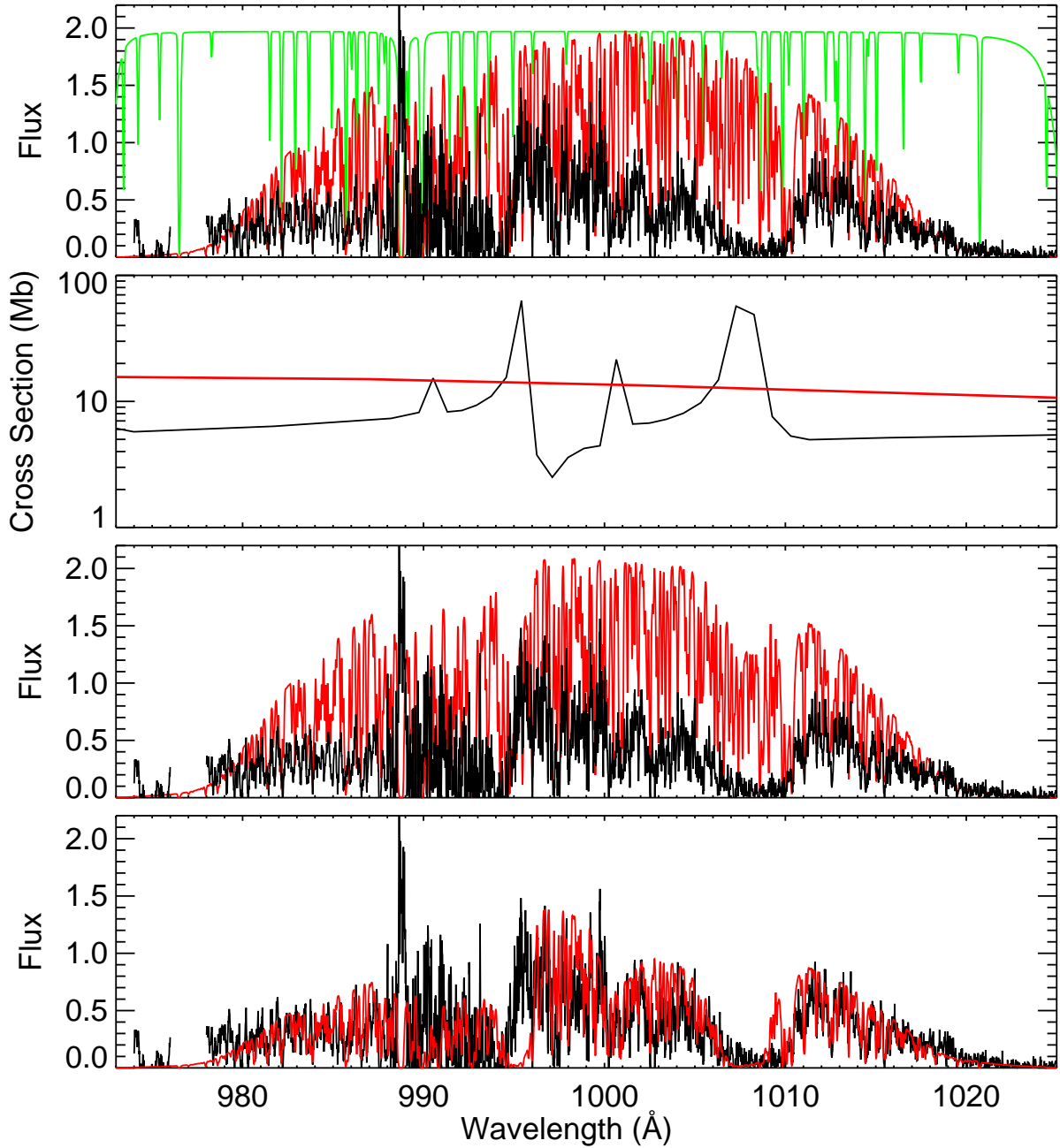}
\caption{\footnotesize First panel: \fuse\/ spectrum of the Bright Star in 47 Tuc (black) overplotted by a synthetic stellar spectrum (red). The normalized curve represents our ISM absorption model (green).  The stellar model assumes a nitrogen abundance [N/H] = -1 and is scaled to match the HUT spectrum between 1400 and 1800 \AA.  The emission feature at 989 \AA\ is geocoronal \oone.  Second panel: Photoionization cross-sections for the first excited state of \none\ (2s$^2$ 2p$^3$ $^2$D$^0$) from the Opacity Project (OP; black).  The red curve represents the resonance-averaged cross sections employed by the atomic models of  Lanz \& Hubeny.  Third panel: We insert the OP cross sections, at full resolution, into the atomic model.  A broad absorption feature appears between 1005 and 1010 \AA. Fourth panel: We increase the nitrogen abundance by a factor of 12.  The general features of the stellar spectrum are now well reproduced.  Note: flux units are $10^{-13}$ \flux.}
\label{fig_fuse}
\end{figure*}

Our initial model is presented in the first panel of \fig{fig_fuse}.  The black curve represents the observed spectrum.  The emission feature at 989 \AA\ is geocoronal \oone.  The red curve represents the synthetic spectrum constructed by SYNSPEC, computed and scaled as described above.  The normalized green spectrum represents interstellar absorption features, which are included in the stellar model.  The model significantly overpredicts the stellar flux in the region between Lyman $\gamma$ (973 \AA) and Lyman $\beta$ (1026 \AA).  Note, in particular, the sharp breaks in the stellar spectrum at 995 and 1010 \AA\ and the broad absorption trough between 1005 and 1010 \AA.  

\begin{figure}
\epsscale{1.32}
\plotone{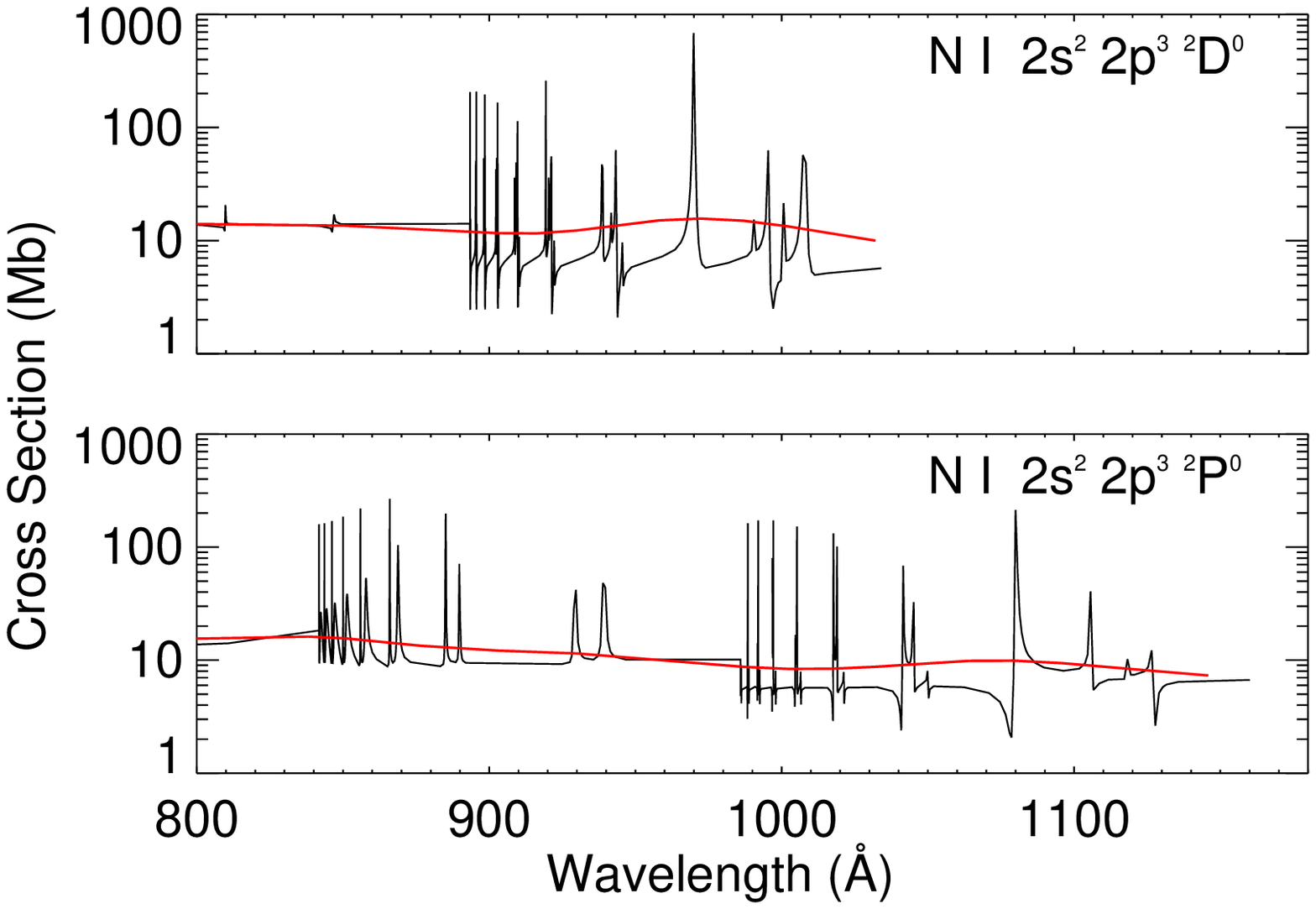}
\caption{Photoionization cross-sections for the first (top panel) and second (bottom panel) excited states of \none.  The black curve represents the cross sections from the Opacity Project at their full resolution.  The red curve represents the resonance-averaged cross sections employed by the atomic models of  Lanz \& Hubeny.}
\label{fig_cross_sections}
\end{figure}

The second panel of \fig{fig_fuse} presents the photoionization cross-sections as a function of wavelength (black curve) for the first excited state of \none\ (2s$^2$ 2p$^3$ $^2$D$^0$) computed by the Opacity Project (OP; see \citealt{Seaton:94} for a general overview).  We see that the sharp breaks and broad troughs in the stellar spectrum correspond to resonances in the photoionization cross sections.  These resonances are more obvious in \fig{fig_cross_sections}, which presents OP cross sections for the first (upper panel) and second (lower panel) excited states of \none.  The \citet{Lanz:Hubeny:2007} atomic models employ OP cross sections, but use resonance-averaged values (red curves in both figures), which cannot reproduce the observed structure.  While the resonances shown in \fig{fig_fuse} are relatively broad, many others are unresolved by the OP calculations \citep{Lanz:Hubeny:ASPC2003}.  Smoothed cross sections are sufficient for modeling the opacity in stellar atmospheres, the purpose for which these atomic models were constructed.

To overcome this shortcoming, we insert the OP cross sections at full resolution into the atomic model for \none\ (using the recipe provided in the SYNSPEC Users Guide) and generate a new synthetic spectrum (third panel of \fig{fig_fuse}).  While the model changes only slightly, we do note the appearance of a broad absorption feature between 1005 and 1010 \AA.  Finally, we increase the nitrogen abundance by a factor of 12 (to 1.2 times the solar value) and plot the result in the fourth panel of \fig{fig_fuse}.  The general features of the stellar spectrum are now well reproduced, but their wavelengths are shifted by a few \AA ngstroms.  This is not surprising; uncertainties of 2 to 3\% in the energies of the OP cross sections can shift the resonances from their observed wavelengths \citep{Lanz:Hubeny:ASPC2003}.   \citet{Lanz:1996} have shown how resonances in the cross sections of \sitwo\ can shift due to small variations in their computed energies.  Between our initial and final models, the integrated flux between 972 and 1026 \AA\ falls by a factor of two.

To reproduce the observed spectrum, the model requires a super-solar nitrogen abundance.  How does this value compare with that derived from bound-bound transitions in the stellar spectrum?  To reduce the impact of non-LTE effects on our analysis \citep{Przybilla:Butler:2001}, we have identified four uncontaminated features due to absorption by the first excited state of \none\ (the same state that is responsible for the resonances discussed above), at 1165.6, 1169.7, 1176.5, and 1176.6 \AA.  They yield N/H ratios between 0.4 and 0.6 times the solar value, about half the value derived from the bound-free continuum.  This result suggests that the \none\ cross sections predicted by the OP calculations are too small by a factor of two.  One caveat: As discussed below, bound-free transitions of \cone\ depress the stellar continuum at wavelengths shorter than 1100 \AA, so uncertainties in the carbon abundance may also be at play.

We cannot use ground-state transitions to constrain the star's photospheric nitrogen abundance, as their spectral features are blended with strong ISM lines.  A nitrogen abundance of half the solar value is not unreasonable for this object.  \citet{Briley:2004} find that N/H ranges from 0.15 to 5 times solar among main-sequence stars in 47~Tuc.  As discussed below, we  expect the nitrogen abundance of a post-AGB star to be twice that of a main-sequence object.

\subsection{An Additional Absorption Feature}\label{sec_otherlines}

\begin{figure}
\epsscale{1.3}
\plotone{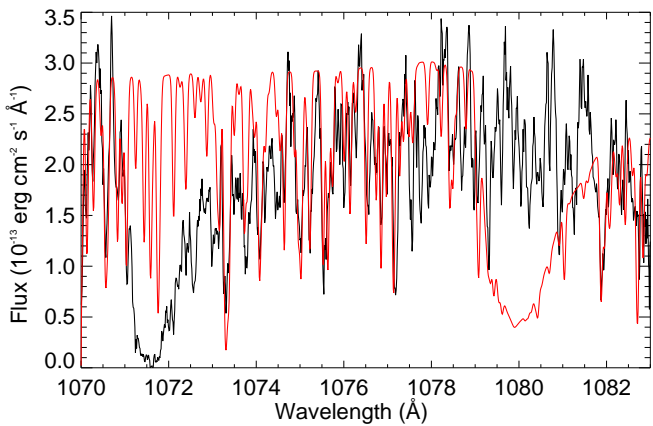}
\caption{\fuse\/ spectrum of the Bright Star in 47 Tuc (black) over-plotted by a synthetic spectrum including OP cross sections at full resolution (red).  The deep absorption feature present in the data at 1072 \AA\ appears in the model at 1080 \AA.  It is caused by a resonance in the bound-free absorption cross sections of the second excited state of \none\ (2s$^2$ 2p$^3$ $^2$P$^0$).}
\label{fig_1072}
\end{figure}

Another feature can be explained by bound-free absorption from the second excited state of \none\ (2s$^2$ 2p$^3$ $^2$P$^0$).  In \fig{fig_1072}, the black curve represents the observed spectrum, while the red curve represents our model.  The model includes OP cross sections for the second excited state of \none, again at full resolution, and assumes a nitrogen abundance N/H of 1.2 times the solar value.  The broad absorption feature seen near 1072 \AA\ in the data is shifted by about 8 \AA\ in the model.  This shift corresponds to an energy error of less than 1\%, well within the uncertainty of the OP calculations.  

\section{Discussion}\label{sec_discussion}

Both the anomalous opacity and the peculiar absorption pattern seen in the FUV spectrum of the Bright Star in 47 Tuc are due to resonances in the photoionization cross sections of \none\ in the first excited state.  Including this opacity in the model, by employing OP cross sections at their full resolution and raising the model's nitrogen abundance, reduces the model flux by a factor of two and reproduces the broad absorption troughs centered at 995 and 1008 \AA.  A second absorption feature centered near 1072 \AA\ can be reproduced by including opacity from \none\ in the second excited state, though the model feature is offset in wavelength.

The importance of photoionization resonances in the UV spectra of late-B through F-type stars has been explored by \citet{Lanz:1996}, who computed bound-free cross sections of \sitwo\ and used them to model resonances in the \iue\/ spectrum of the Ap Si star HD~34452.  Photoionization resonances have been identified in the EUV spectrum of the white dwarf GD~246 \citep{Vennes:93} and the X-ray spectra of several X-ray binaries \citep{Juett:2004, Garcia:2011, Gatuzz:2013}.

As a simple test of the importance of nitrogen opacity in Population I giants, we generated a synthetic spectrum using a model atmosphere with the same temperature and gravity as the Bright Star, but with solar metallicity.  The spectrum shows almost no flux at wavelengths shorter than 1100 \AA; the continuum is obliterated by bound-free absorption from ground-state atomic carbon.  Only when the carbon abundance is reduced to a fraction of the solar value is the FUV continuum bright enough to reveal nitrogen resonance absorption.

According to stellar evolutionary theory, a star's arrival on the red-giant branch is accompanied by a deepening of its convective envelope, which brings the ashes of hydrogen burning to the stellar surface.  This process, known as first dredge-up, doubles the surface $^{14}$N abundance, reduces the $^{12}$C abundance by about 30\%, and leaves the abundance of $^{16}$O unchanged \citep{Iben:Renzini:83}.  No further mixing is predicted for the low-mass stars present in globular clusters today.

The Bright Star in 47~Tuc may be the ideal candidate to exhibit strong resonances of atomic nitrogen.  As a post-AGB star in a globular cluster, it combines the high nitrogen abundance necessary to carve these features in its FUV spectrum with the low carbon abundance necessary to reveal them.  Population~I stars will not show these resonances unless their carbon abundances are considerably less than the solar value.
 
\acknowledgments

Portions of this work were supported by NASA Grant \#NNX06AD30G to the Johns Hopkins University. Its publication is supported by the STScI Director's Discretionary Research Fund.  This work has made use of NASA's Astrophysics Data System Bibliographic Services (ADS) and the Mikulski Archive for Space Telescopes (MAST), hosted at the Space Telescope Science Institute. STScI is operated by the Association of Universities for Research in Astronomy, Inc., under NASA contract NAS5-26555. Support for MAST for non-\hst\ data is provided by the NASA Office of Space Science via grant NAG5-7584 and by other grants and contracts.  IDL is a registered trademark of Exelis Visual Information Solutions, Inc., for its Interactive Data Language software.  


{\it Facilities:} \facility{FUSE}.




\begin{thebibliography}{25}
\expandafter\ifx\csname natexlab\endcsname\relax\def\natexlab#1{#1}\fi

\bibitem[{{Alves-Brito} {et~al.}(2005)}]{Alves-Brito:05}
{Alves-Brito}, A. {et~al.} 2005, \aap, 435, 657

\bibitem[{{Briley} {et~al.}(2004){Briley}, {Harbeck}, {Smith}, \&
  {Grebel}}]{Briley:2004}
{Briley}, M.~M., {Harbeck}, D., {Smith}, G.~H., \& {Grebel}, E.~K. 2004, \aj,
  127, 1588

\bibitem[{{Castelli} \& {Kurucz}(2003)}]{Castelli:Kurucz:03}
{Castelli}, F. \& {Kurucz}, R.~L. 2003, in IAU Symp. 210, Modelling of Stellar
  Atmospheres, ed. N.~{Piskunov}, W.~W. {Weiss}, \& D.~F. {Gray} (San
  Francisco: ASP), A20

\bibitem[{{Davidsen} {et~al.}(1992)}]{HUT1CAL1}
{Davidsen}, A.~F. {et~al.} 1992, \apj, 392, 264

\bibitem[{{de Boer}(1985)}]{deBoer:85}
{de Boer}, K.~S. 1985, \aap, 142, 321

\bibitem[{{Dixon} {et~al.}(1995){Dixon}, {Davidsen}, \& {Ferguson}}]{Dixon:95}
{Dixon}, W.~V., {Davidsen}, A.~F., \& {Ferguson}, H.~C. 1995, \apjl, 454, L47

\bibitem[{{Dixon} {et~al.}(2007)}]{Dixon:07}
{Dixon}, W.~V. {et~al.} 2007, \pasp, 119, 527

\bibitem[{{Feast} \& {Thackeray}(1960)}]{Feast:Thackeray:60}
{Feast}, M.~W. \& {Thackeray}, A.~D. 1960, \mnras, 120, 463

\bibitem[{{Garc{\'{\i}}a} {et~al.}(2011){Garc{\'{\i}}a}, {Ram{\'{\i}}rez},
  {Kallman}, {Witthoeft}, {Bautista}, {Mendoza}, {Palmeri}, \&
  {Quinet}}]{Garcia:2011}
{Garc{\'{\i}}a}, J., {Ram{\'{\i}}rez}, J.~M., {Kallman}, T.~R., {Witthoeft},
  M., {Bautista}, M.~A., {Mendoza}, C., {Palmeri}, P., \& {Quinet}, P. 2011,
  \apjl, 731, L15

\bibitem[{{Gatuzz} {et~al.}(2013)}]{Gatuzz:2013}
{Gatuzz}, E. {et~al.} 2013, \apj, 768, 60

\bibitem[{{Harris}(1996)}]{Harris:96}
{Harris}, W.~E. 1996, \aj, 112, 1487

\bibitem[{{Heiles} \& {Cleary}(1979)}]{Heiles:79}
{Heiles}, C. \& {Cleary}, M.~N. 1979, Australian J. Phys. Astrophys. Suppl.,
  47, 1

\bibitem[{{Hubeny}(1988)}]{Hubeny:88}
{Hubeny}, I. 1988, Comput. Phys. Comm., 52, 103

\bibitem[{{Iben} \& {Renzini}(1983)}]{Iben:Renzini:83}
{Iben}, Jr., I. \& {Renzini}, A. 1983, \araa, 21, 271

\bibitem[{{Juett} {et~al.}(2004){Juett}, {Schulz}, \&
  {Chakrabarty}}]{Juett:2004}
{Juett}, A.~M., {Schulz}, N.~S., \& {Chakrabarty}, D. 2004, \apj, 612, 308

\bibitem[{{Kurucz}(1992)}]{Kurucz:92}
{Kurucz}, R.~L. 1992, in IAU Symp. 149, The Stellar Populations of Galaxies,
  ed. B.~{Barbuy} \& A.~{Renzini} (Dordrecht: Kluwer), 225

\bibitem[{{Lanz} {et~al.}(1996){Lanz}, {Artru}, {Le Dourneuf}, \&
  {Hubeny}}]{Lanz:1996}
{Lanz}, T., {Artru}, M.-C., {Le Dourneuf}, M., \& {Hubeny}, I. 1996, \aap, 309,
  218

\bibitem[{{Lanz} \& {Hubeny}(2003)}]{Lanz:Hubeny:ASPC2003}
{Lanz}, T. \& {Hubeny}, I. 2003, in ASP Conf. Ser. 288, Stellar Atmosphere
  Modeling, ed. I.~{Hubeny}, D.~{Mihalas}, \& K.~{Werner} (San Francisco: ASP),
  117

\bibitem[{{Lanz} \& {Hubeny}(2007)}]{Lanz:Hubeny:2007}
{Lanz}, T. \& {Hubeny}, I. 2007, \apjs, 169, 83

\bibitem[{{Moos} {et~al.}(2000)}]{Moos:00}
{Moos}, H.~W. {et~al.} 2000, \apjl, 538, L1

\bibitem[{{Przybilla} \& {Butler}(2001)}]{Przybilla:Butler:2001}
{Przybilla}, N. \& {Butler}, K. 2001, \aap, 379, 955

\bibitem[{{Sahnow} {et~al.}(2000)}]{Sahnow:00}
{Sahnow}, D.~J. {et~al.} 2000, \apjl, 538, L7

\bibitem[{{Seaton} {et~al.}(1994){Seaton}, {Yan}, {Mihalas}, \&
  {Pradhan}}]{Seaton:94}
{Seaton}, M.~J., {Yan}, Y., {Mihalas}, D., \& {Pradhan}, A.~K. 1994, \mnras,
  266, 805

\bibitem[{{Stecher} {et~al.}(1997)}]{Stecher:97}
{Stecher}, T.~P. {et~al.} 1997, \pasp, 109, 584

\bibitem[{{Vennes} {et~al.}(1993){Vennes}, {Dupuis}, {Rumph}, {Drake},
  {Bowyer}, {Chayer}, \& {Fontaine}}]{Vennes:93}
{Vennes}, S., {Dupuis}, J., {Rumph}, T., {Drake}, J., {Bowyer}, S., {Chayer},
  P., \& {Fontaine}, G. 1993, \apjl, 410, L119

\end{thebibliography}
\end{document}